\documentclass[11pt,american,twoside,a4wide]{article}
\usepackage[T1]{fontenc}
\usepackage[latin1]{inputenc}
\usepackage{latexsym}
\usepackage{amsmath}
\usepackage{bbm}
\usepackage{amssymb}
\usepackage{babel}
\usepackage{amsfonts}
\usepackage{times}
\usepackage{theorem}
\usepackage{epsfig}
\usepackage{enumerate}
\usepackage{color}
\usepackage[active]{srcltx}
\usepackage[colorlinks=true]{hyperref}
\usepackage{fancyhdr}

\numberwithin{equation}{section}

\newcounter{smallarabics}

\newcounter{smallroman}

\newcommand{\ben}{\begin{enumerate}[{\rm (1)}]}
\newcommand{\een}{\end{enumerate}}

\newtheorem{theoreme}{Theorem}[section]
\newtheorem{proposition}[theoreme]{Proposition}
\newtheorem{lemma}[theoreme]{Lemma}
\newtheorem{definition}[theoreme]{Definition}

\def\rr{{\mathbb R}}
\def\zz{{\mathbb Z}}
\def\cc{{\mathbb C}}

\def\Z{{\mathbb Z}}

\def\textsl{{}}

\def\Im{{\rm Im}\,}

\def\c0inf{C_0^\infty}
\def\bep{\begin{proposition}}
\def\eep{\end{proposition}}

\def\proof{\noindent {\bf Proof.}\ \ }

\def\cH{{\cal  H}}

\newcommand{\bra}{\langle} 
\newcommand{\ket}{\rangle}

\def\per{{\rm per}}

\def\i{{\rm i}}
\newcommand{\beq}{\begin{equation}}
\newcommand{\eeq}{\end{equation}}
\newcommand{\bear}[1]{\begin{array}{#1}}
\newcommand{\ear}{\end{array}}

\def\sp{{\hat e}}

\newcommand{\e}{\mathrm{e}}
\renewcommand{\i}{\mathrm{i}}

\renewcommand{\d}{\mathrm{d}}

\def\qed{$\Box$\medskip}

\def\cJ{{\cal J}}

\def\bel{\begin{lemma}}
\def\eel{\end{lemma}}
\def\bet{\begin{theoreme}}
\def\eet{\end{theoreme}}
\def\bed{\begin{definition}}
\def\eed{\end{definition}}
\def\bar{\overline}

\def\12{\frac{1}{2}}

\def\e{{\rm e}}
\def\cD{{\cal D}}

\def\d{{\rm d}}
\def\Ran{{\rm Ran}\,}

\def\one{{\mathbbm 1}}

\def\cH{{\cal H}}

\def\ac{{\rm ac}}

\def\sp{{\rm sp}}

\def\tr{{\rm tr}}

\def\bra{\langle}
\def\ket{\rangle}

\newcommand{\ds}{\displaystyle}
\begin{document}
\def\today{}
\title{Conductance and absolutely continuous spectrum of 1D samples}
\author{L. Bruneau$^{1}$, V. Jak\v{s}i\'c$^{2}$, Y. Last$^3$, C.-A. Pillet$^4$
\\ \\ 
$^1$ D\'epartement de Math\'ematiques and UMR 8088\\
CNRS and Universit\'e de Cergy-Pontoise\\
95000 Cergy-Pontoise, France
\\ \\
$^2$Department of Mathematics and Statistics\\ 
McGill University\\
805 Sherbrooke Street West \\
Montreal,  QC,  H3A 2K6, Canada
\\ \\
$^3$Institute of Mathematics\\
The Hebrew University\\
91904 Jerusalem, Israel
\\ \\
$^4$Aix-Marseille Universit\'e, CPT, 13288 Marseille cedex 9, France\\
CNRS, UMR 7332, 13288 Marseille cedex 9, France\\
Universit\'e de Toulon, CPT, B.P. 20132, 83957 La Garde cedex, France\\
FRUMAM
}
\maketitle
\thispagestyle{empty}
\begin{quote}
\noindent{\bf Abstract.} We characterize   the absolutely continuous spectrum of the  one-dimensional Schr\"odinger operators $h=-\Delta+v$ acting on  $\ell^2(\zz_+)$ in terms  of   the  limiting behaviour  of the Landauer-B\"uttiker  and Thouless conductances of the associated finite samples. 
The  finite sample is defined  by  restricting  $h$ to  a  finite interval $[1, L]\cap\zz_+$ and the conductance refers to the  charge current  across the 
sample in  the open quantum system obtained by  attaching independent electronic reservoirs to the sample ends. 
Our main result is  that the conductances associated to an energy interval $I$ are  non-vanishing in the limit $L\to \infty$ 
 iff $\sp_{\rm ac}(h)\cap I \neq \emptyset$.   We also discuss the relationship between this result and the  Schr\"odinger Conjecture 
\cite{Av, BJP}.
\end{quote}
\section{Introduction}\label{sec:intro}

This paper  concerns a  connection between  two directions  of research: 
transport theory of open quantum systems and spectral theory of discrete  
Schr\"odinger operators. The simplest open quantum system where this connection  is 
exhibited, the so-called electronic black box model (EBBM), consists of 
a finite sample connecting two free electron reservoirs. The model is 
considered in the independent electron and tight binding approximations and 
the  object of study is the charge current across the  sample induced by the 
voltage differential between the reservoirs.
The celebrated Landauer-B\"uttiker and Thouless current/conductance formulas 
of finite samples arose from such considerations. 

\begin{figure}
\centering
\includegraphics[scale=0.5]{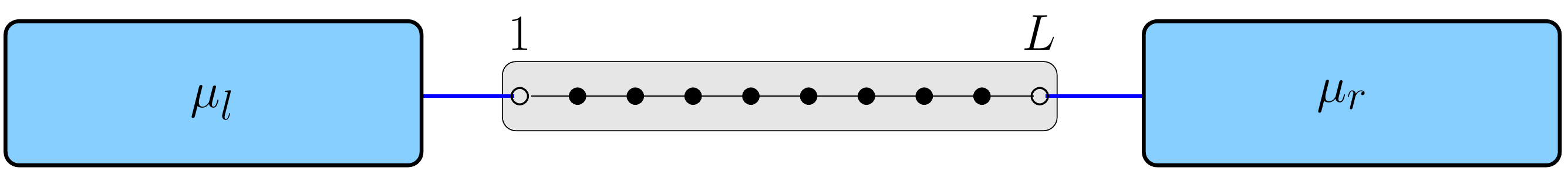}
\caption{A finite sample of length $L$ coupled to two electronic reservoirs}
\label{Fig3}
\end{figure}

In this work we shall restrict ourselves to 1D geometry. The one-particle 
configuration space of a sample of length $L$ is the finite set 
$\zz_L=\{1,2,\ldots,L\}$. Left and right electronic reservoirs are attached 
to the sample at site $1$ and $L$, respectively (see Figure~\ref{Fig3}).
We denote by $\zz_+$ the positive integers. To a potential 
$v:\zz_+\rightarrow\rr$ we associate the discrete Schr\"odinger operator
\[
h=-\Delta + v,
\]
acting on the Hilbert space $\ell^2(\zz_+)$\footnote{For our purposes, the 
choice of boundary condition is irrelevant and for definiteness we will 
use Dirichlet b.c.}. 
We shall view $\zz_+$  as the one-particle configuration space and $h$ as the 
Hamiltonian of the extended sample. The  one-particle Hamiltonian of the sample of length $L$ is obtained by restricting $h$ to  $\ell^2(\zz_L)$.
We are interested in the relationship between the spectral properties of the
extended Hamiltonian $h$ and the limiting values of the Landauer-B\"uttiker 
and Thouless current/conductance of the finite sample as 
$L\rightarrow \infty$. More specifically, we will focus on the relationship 
between: 

\begin{enumerate}[{\rm (A)}]
\item The physical characterization of the conducting  regime of the extended 
sample as the set of energies at which the current/conductance is non-vanishing in the limit $L\rightarrow\infty$.
\item The mathematical characterization of the conducting regime of the extended sample as the absolutely 
continuous spectrum of $h$, denoted $\sp_\ac(h)$.
\end{enumerate}

The recent rigorous proofs of the Landauer-B\"uttiker and Thouless 
current/conductance formulas from the first principles of quantum statistical
mechanics~\cite{AJPP,N,BJLP,BSP} have opened the way to the study of the 
equivalence ${\rm (A)}\Leftrightarrow {\rm (B)}$. Some preliminaries are
required to formulate this equivalence in mathematically precise terms.

We shall assume that the left and right reservoirs are in thermal equilibrium 
at zero temperature and chemical potentials $\mu_l<\mu_r$. The role of the 
chemical  potentials is to "probe" the sample in the  energy interval 
$[\mu_l,\mu_r]$. In the large time limit, the potential differential 
$\mu_r-\mu_l$ induces a steady charge current across the sample. The
expectation value $\cJ_{\rm LB}(L, \mu_l, \mu_r)$ of this steady current is 
given by the Landauer-B\"uttiker
formula~\eqref{eq:lbformula}-\eqref{eq:transmissionproba}. This formula 
depends intrinsically on the structure of the reservoirs and on the form of 
their coupling to the sample. One particular choice of the 
reservoirs/couplings leads to the Thouless current formula which we denote 
by $\cJ_{\rm Th}(L, \mu_l, \mu_r)$, see 
Section~\ref{ssec:thoulessconductance}. The respective conductances are  
\[
G_{{\rm LB}}(L, \mu_l, \mu_r)
=\frac{1}{\mu_r-\mu_l}\cJ_{{\rm LB}}(L, \mu_l, \mu_r) \quad 
{\rm and} \quad 
G_{{\rm Th}}(L, \mu_l, \mu_r)
=\frac{1}{\mu_r-\mu_l}\cJ_{{\rm Th}}(L, \mu_l, \mu_r).
\]
In our analysis, current and conductance play similar roles, and in the sequel 
we  will switch between these two notions depending on notational 
convenience. We shall review the Landauer-B\"uttiker and Thouless formulas in 
Section~\ref{sec-cond}. To avoid trivialities when using the
Landauer-B\"uttiker formula we shall assume that the reservoirs are 
transparent for the energies in the interval 
$(\mu_l,\mu_r)$  (see Definition~\ref{transparent} below).

A mathematically precise formulation of the equivalence 
${\rm (A)}\Leftrightarrow{\rm (B)}$ is the object of the following two 
conjectures, which should hold for any potential $v$: 

\begin{quote}{\bf Conjecture I.} If $(\mu_l, \mu_r)\cap \sp_\ac(h)=\emptyset$, then 
\[\lim_{L\rightarrow \infty} G_{\#}(L, \mu_l, \mu_r)=0,\]
where $\#$ stands for LB or   Th. 
\end{quote}

\begin{quote}{\bf Conjecture II.} If $(\mu_l, \mu_r)\cap \sp_\ac(h)\not=\emptyset$, then 
\[\liminf_{L\rightarrow \infty} G_{\#}(L, \mu_l, \mu_r)>0,\]
where $\#$ stands for LB or   Th. 
\end{quote}

Just like the celebrated Schr\"odinger Conjecture~\cite{MMG, Si1, Av}, which we 
will  discuss below, Conjectures I and II are rooted in the formal 
computations and implicit assumptions of the physicists working on the 
subject. To the best of our knowledge, they 
were first 
formulated in the above mathematical form in~\cite{La} which treats  the case $\#={\rm Th}$  in the setting of
ergodic Schr\"odinger operators. We refer the reader to~\cite{La} for 
references regarding early physicists' work that motivated the conjectures 
and to~\cite{CGM} for supporting numerical results. Conjectures I and II
are also of importance for the foundations of quantum mechanics 
since they would provide the 
first complete dynamical characterization of the absolutely continuous 
spectrum of Schr\"odinger operators.\footnote{The Landauer-B\"uttiker and 
Thouless conductance formulas~\cite{AJPP,N,BJLP,BSP} concern
the steady state value reached by the charge current in the large time limit 
and  hence have a dynamical origin; see~\cite{BJLP2} for a discussion of 
this point in the context of spectral theory.} 

A strong form of Conjectures I and II in the case $\#={\rm LB}$  was studied 
in the recent work~\cite{BJP}. There, the focus was on the
Landauer-B\"uttiker spectral density defined by 
\beq
{\cal D}_{\rm LB}(L, E)
=\lim_{\delta \downarrow 0}G_{\rm LB}(L, E-\delta, E+\delta).
\label{density}
\eeq
The limit~\eqref{density} exists for Lebesgue a.e.\;$E\in \rr$, takes values 
in $[0,(2\pi)^{-1}]$, and is such that
\beq
\cJ_{{\rm LB}}(L, \mu_l, \mu_r)
=\int_{\mu_l}^{\mu_r}{\cal D}_{\rm LB}(L, E)\d E.
\label{eq:integrateddensity}
\eeq
Although the density ${\cal D}_{\rm LB}(L, E)$ depends intrinsically on the structure of the reservoirs and the choice 
 of the coupling, it does not depend on the choice of the thermodynamical states of the reservoirs, and in 
particular it does not depend on the choice of $\mu_{l/r}$. For more information about ${\cal D}_{\rm LB}(L, E)$, we refer the reader to 
Section~\ref{ssec:lbconductance}.  

In our setting, the transfer matrices of $h$ provide the link between transport and spectrum. We denote by 
\beq
T(L, E)
=\left[ \begin{matrix} v(L)-E & -1 \\ 1 & 0 \end{matrix} \right]\cdots \left[\begin{matrix} v(1)-E & -1 \\ 1 & 0 \end{matrix} \right]
\label{transfer}
\eeq
the transfer matrix of $h$ between the sites $1$ and $L$ at energy $E$.  It is easily shown that
\beq
T(L, E)
=\left[ \begin{matrix} u_D(L+1, E) & u_N(L+1, E)\\ 
u_D(L, E) & u_N(L, E) \end{matrix}  \right],
\label{transfer-1}
\eeq
where $u_X(L, E)$, $X\in \{D, N\}$, is the unique solution of the 
Schr\"odinger equation $hu=Eu$ with the boundary condition $u(1)=1$, $u(0)=0$  
in the case $X=D$, and the boundary condition $u(1)=0$, $u(0)=1$ in the case
$X=N$. In~\cite{LS} it was proven  that 
\beq
\Sigma_\ac=\left\{ E\,:\, \liminf_{L\rightarrow \infty}
\frac{1}{L}\sum_{\ell=1}^L\|T(\ell, E)\|^2<\infty\right\},
\label{LS-1}
\eeq
where $\Sigma_\ac$  is  the essential support of the absolutely continuous 
spectrum of $h$  and the equality is modulo a set  of Lebesgue measure 
zero.\footnote{In the sequel, whenever the meaning is clear within the 
context, we shall write $S_1=S_2$ for two subsets of $\rr$ if the Lebesgue 
measure of their symmetric difference is equal to zero. Similarly, 
we shall write $S_1\subset S_2$ if the Lebesgue measure of $S_2\setminus S_1$ is zero, etc.} Let 
\[
{\mathfrak S}_0=\{E\,:\,\sup_L\|T(L,E)\|<\infty\}, \qquad 
{\mathfrak S}_1=\{E\,:\,\liminf_{L\rightarrow\infty}\|T(L,E)\|<\infty\}.
\]
It follows from~\eqref{LS-1} that 
\beq 
{\mathfrak S}_0\subset \Sigma_\ac\subset {\mathfrak S}_1.
\label{flight}
\eeq
We remark that the first inclusion goes back to~\cite{GP} (see 
also~\cite{Si0}), while the second has a direct proof which 
we will sketch in Remark~6 after Theorem~\ref{main-th}. If the equality 
\beq 
{\mathfrak S}_0=\Sigma_\ac={\mathfrak S}_1
\label{schr}
\eeq
holds, one says that the operator $h$ has the {\em Schr\"odinger Property}.

The main result of \cite{BJP} links the sets ${\mathfrak S}_0$ and 
${\mathfrak S}_1$ to the LB conductance as follows: 
\beq
\{E\,:\, \liminf_{L\rightarrow \infty} {\cal D}_{\rm LB}(L, E)>0\}={\mathfrak S}_0, \qquad \{E\,:\, \limsup_{L\rightarrow \infty} {\cal D}_{\rm LB}(L, E)>0\}={\mathfrak S}_1.
\label{relat}
\eeq
An easy application of Fatou's Lemma and Lebesgue's dominated convergence 
theorem  shows that    these relations and  the Schr\"odinger Property imply 
Conjectures I and II for the LB conductance. 
From the physical point of view,  the Schr\"odinger Property  is also a  
strengthening of the LB part of the  Conjectures I and II  due to the role the 
density  ${\cal D}_{\rm LB}(L, E)$ plays in linear response theory and 
fluctuation-dissipation theorem (see~\cite{JOPP, BJLP2} for a pedagogical 
discussion  of this topic).


At the time of the completion of the work~\cite{BJP}, it was generally  
believed that any half-line discrete Schr\"odinger operator has the  
Schr\"odinger Property, a fact known as the {\em Schr\"odinger Conjecture}.   
From the mathematical point of view, for many years the Schr\"odinger 
Conjecture was arguably the single most important open problem in  general 
spectral theory of Schr\"odinger operators. The  main goal of the 
work~\cite{BJP} was to point   out that  the Schr\"odinger Conjecture is 
closely linked to the  LB conductance and that it can be viewed as a 
{\em strong version} of the LB part of the Conjectures I and II. 

Spectacularly, in the recent  work~\cite{Av}, Avila has constructed a 
counterexample to the Schr\"odinger Conjecture. Even more strikingly, this 
counterexample is in the context of ergodic Schr\"odinger operators for which 
$\Sigma_\ac$ has a very rigid structure dictated by the  Kotani Theory. 
In the ergodic setting, $v_\omega(n)=V(S^n\omega)$ where $\Omega$ is a 
measure space, $V:\Omega\to\rr$ is a bounded 
measurable map, and $S$ is an ergodic invertible transformation of $\Omega$. 
The Lyapunov exponent of the model is 
\beq 
\gamma(E)=\lim_{L\rightarrow \infty}\frac{1}{L}\log \|T_\omega(L, E)\|,
\label{Kotani}
\eeq
where, for given $E$, the limit exists for a.e. $\omega$ and does not depend 
on $\omega$. The Kotani Theory~\cite{Ko,Si2,DS} gives
\beq 
\Sigma_\ac=\{E\,:\, \gamma(E)=0\}.
\label{kot}
\eeq
This characterization of $\Sigma_\ac$ and the second inclusion 
in~\eqref{flight} imply that in the  ergodic 
setting one always has $\Sigma_\ac={\mathfrak S}_1$ with probability one. 
We also mention the result of Deift and Simon~\cite{DS}, which gives that with 
probability one (compare with~\eqref{LS-1})
\beq
\Sigma_\ac=\left\{ E\,:\, \limsup_{L\rightarrow \infty}
\frac{1}{L}\sum_{\ell=1}^L\|T_\omega(\ell, E)\|^2<\infty\right\}.
\label{DS-1}
\eeq
Avila~\cite{Av} constructs $\Omega$, $V$,  and an (uniquely) ergodic 
transformation $S$  such that there is a set $\Lambda\subset\Sigma_\ac$ of 
positive Lebesgue measure with the property that for any $E\in\Lambda$ and 
a.e.\;$\omega\in\Omega$ any non-trivial (generalized) eigenfunction of 
$h_\omega$ is unbounded and hence so is $\|T_\omega(L,E)\|$. In other words,  
for a set of $\omega$'s of probability one the Lebesgue measure  of 
$\Sigma_\ac\setminus {\mathfrak S}_0$ is strictly positive.

The dramatic failure of the Schr\"odinger Conjecture, or, equivalently, of  
the {\em strong version} of the Conjectures I and II, does  not exclude the 
possibility that these conjectures hold in their original form.
The main goal of our work is to address this point. In view of Avila's 
counterexample, it is important  to distinguish between the ergodic and the 
deterministic case.

In the ergodic setting and the LB case, the validity of Conjectures~I and 
II  follows from~\eqref{Kotani}  and the results of~\cite{BJP} (\cite{BJ}, 
see~\cite{BJLP2} for a pedagogical discussion). In the ergodic 
setting and the Th case, the conjectures were proven in the unpublished part 
of~\cite{La}. The special aspect of the ergodic setting is that the energy 
averaging leads to a priori estimates on the size of transfer 
matrices\footnote{This estimates are deterministic in nature;  see Remark~6 
after Theorem~\ref{main-th}.} that can be effectively combined with Kotani
Theory  to prove Conjectures~I and II. In turn, these results are one of the 
reasons why Avila's counterexample is so surprising: in the ergodic setting 
the averaged forms of the Schr\"odinger Conjecture were known to hold in the mathematical sense  
(relation~\eqref{DS-1}) and the physical sense (Conjectures~I and II). 
We refer the reader to the Introduction in~\cite{Av} for an 
additional  discussion of this point. 

This leaves us with the  deterministic case where, unlike in the ergodic 
case,  the validity of Conjectures I and II for all potentials $v$ was far 
from clear. Our main result settles this case.

\bet\label{main-th} 
For any potential $v$ on $\zz_+$, any  $\mu_l <\mu_r$, and any sequence  $(L_k)$ of positive integers satisfying $\lim L_k=\infty$,  the following statements are equivalent: 
\begin{enumerate}[{\rm (1)}]
\item $$(\mu_l, \mu_r)\cap \sp_\ac(h)=\emptyset.$$
\item 
\[\lim_{k\rightarrow \infty}\int_{\mu_l}^{\mu_r}\|T(L_k, E)\|^{-2}\d E =0.
\]
\item 
\[\lim_{k\rightarrow \infty}G_{\rm LB}(L_k, \mu_l, \mu_r)=0.
\]
\item 
\[\lim_{k\rightarrow \infty}G_{\rm Th}(L_k, \mu_l, \mu_r)=0.
\]
\end{enumerate}
\eet
The equivalences between (1), (3) and (4) correspond exactly to the validity 
of Conjectures~I and II, i.e.\;to the equivalence 
${\rm (A)}\Leftrightarrow {\rm (B)}$.

{\bf Remark 1.} The proof of the implication $(3)\Rightarrow (2)$ requires the 
non-triviality assumption that the reservoirs are transparent for 
the energies in the interval $(\mu_l,\mu_r)$. The precise formulation of this 
assumption is given in Definition~\ref{transparent}.

{\bf Remark 2.} The relevance of $(2)$ in our context stems from~\cite{BJP} 
and, more  implicitly,  from  the  early physicists' works on the subject.
Our  proof of Theorem \ref{main-th} proceeds by establishing the equivalences 
$(2)\Leftrightarrow (1)$, $(2)\Leftrightarrow (3)$, 
$(2)\Leftrightarrow (4)$. 

{\bf Remark 3.} Theorem~\ref{main-th} can be extended  to the case where the sample 
Hamiltonian $h$ is a general half-line Jacobi matrix. In turn, this extension 
allows one  to prove a suitable analog of Theorem~\ref{main-th} in the setting 
where the extended sample is described by an arbitrary Hilbert space and 
Hamiltonian. These extensions  are discussed in the forthcoming review 
article~\cite{BJLP2}.

{\bf Remark 4.} A natural link between the Landauer-B\"uttiker and Thouless 
conductances is provided by the Crystaline Landauer-B\"uttiker conductance 
introduced in~\cite{BJLP}. This conductance has an additional mathematical and 
physical structure that goes beyond Conjectures~I and II and that  
may shed a light on the transport origin of the fundamental results of 
Kotani~\cite{Ko, Si2} and Remling~\cite{Re}. This topic remains to be studied in the future.

{\bf Remark 5.} To the best of our knowledge, the first mathematical results 
regarding the relation between absolutely continuous spectrum 
and conductance go back to~\cite{La}. These results preceded the rigorous 
proofs of the conductance formulas and remained unpublished. The equivalence 
$(1)\Leftrightarrow(4)$ was proven in~\cite{La} in the ergodic setting.  
In  Remark~7  we will comment more on the relation between our work 
and~\cite{La}. 


{\bf Remark 6.} The proofs of the equivalences 
$(1)\Leftrightarrow (2)\Leftrightarrow (3)$ are based on  three ingredients. 
The first ingredient is the second inclusion  in~\eqref{flight},  which is 
proven in~\cite{LS}. We sketch the argument since it sheds some light on the 
mathematical structure behind  the above  equivalences. Let $\nu_X$ be the 
spectral measure for $h$ with Dirichlet $X=D$ or Neumann $X=N$ boundary 
condition. The spectral theorem gives that for all $L$, 
\begin{equation}
\int_\rr |u_X(L, E)|^2 \d\nu_X(E)=1,
\label{air-france}
\end{equation}
where $u_X$, $X\in\{D,N\}$, is defined in~\eqref{transfer-1}. Setting 
\[
\bar\nu(S)=\inf_{{A, B}\atop {S\subset A\cup B}}(\nu_D(A) + \nu_N(B)),
\]
one easily shows that $\bar\nu$ is a Borel measure whose absolutely continuous 
part $\bar \nu_\ac$ is equivalent to $\nu_{X,\ac}$. In particular, 
\[
\Sigma_\ac
=\left\{E\ :\ \frac{\d\bar \nu_{\ac}}{\d E}(E)>0\right\}
=\left\{E\ :\ \frac{\d\nu_{X,\ac}}{\d E}(E)>0\right\}.\]
Relations~\eqref{transfer-1} and~\eqref{air-france} give 
\[
\int_\rr \|T(L, E)\|^2 \d\bar \nu_{\ac}(E)\leq 4, 
\]
and Fatou's Lemma yields
\beq
\Sigma_\ac \subset 
\left\{E: \liminf_{k\rightarrow \infty}\|T(L_k, E)\|<\infty\right\}.
\label{b-class}
\eeq
For details of the arguments we refer the reader to~\cite{LS}. The above 
sketch gives the direct proof of the second inclusion in~\eqref{flight}.  
The relation~\eqref{b-class} yields the implication $(2)\Rightarrow (1)$.  

The  second ingredient is the main technical result of 
\cite{BJP} which gives 
\[
\left\{E\in \Sigma_l\cap \Sigma_r\,:\, 
\lim_{k\rightarrow \infty}{\cal D}_{\rm LB}(L_k, E)=0\right\}
=\left\{E\in \Sigma_l\cap \Sigma_r\,:\, 
\lim_{k\rightarrow \infty}\|T(L_k, E)\|=\infty\right\},
\label{b-1-class}
\]
where $\Sigma_{l/r}$ denotes the essential support of the absolutely 
continuous spectrum of the $l/r$ reservoir (see Eq.~\eqref{eq:Simgalr}).
This relation yields the equivalence  $(2)\Leftrightarrow (3)$.\footnote{One 
can actually prove that 
$C\int_{\mu_l}^{\mu_r}\|T(L, E)\|^{-2}\d E \leq G_{\rm LB}(L, \mu_l, \mu_r) 
\leq C'\int_{\mu_l}^{\mu_r}\|T(L, E)\|^{-2}\d E$ for some constants $C,C'>0$ 
and any $L$; see~\cite{BJLP2}.}

In the ergodic setting the implication $(1)\Rightarrow(2)$ is an immediate 
consequence of the Kotani result~\eqref{kot}. Its proof in the deterministic 
setting  relies on a subtle and surprising result of~\cite{Ca,KR,Si3} which 
is the third ingredient. This result states that if 
$u=(1,0)^T$, then\footnote{We choose 
Dirichlet b.c., although an analogous result holds for any other b.c.}
\[
\frac{1}{\pi}\|T(L, E)u\|^{-2}\d E\rightarrow \d\nu_D(E)
\]
weakly as $L\rightarrow \infty$. 

The details of the proofs are given in Sections~\ref{sec-proof12} 
and~\ref{sec-proof23}. Given the above three ingredients, they are 
surprisingly simple.

{\bf Remark 7.}  Our  proof of the equivalence $(2)\Leftrightarrow(4)$ is 
guided by the results of~\cite{La}. The arguments in~\cite{La} can be 
separated into two parts. The arguments in the first part are 
deterministic in nature and are presented in~\cite{La} in the ergodic setting 
only for  notational convenience. The arguments in the second part  
rely essentially on Kotani Theory and are  applicable only in the ergodic 
setting. In Section~\ref{sec-proof24} we review the deterministic part and 
give novel arguments replacing the ergodic part to complete the 
proof of the equivalence  $(2)\Leftrightarrow(4)$. 

Perhaps the  most interesting consequence of the new arguments concerns periodic 
approximations. The proof of the implication $(1)\Rightarrow (4)$ in~\cite{La} 
is based on the following result of~\cite{La3}. Let 
$h_\omega=-\Delta + v_\omega(n)$, $v_\omega(n)=V(S^n\omega)$, be a full line 
ergodic Schr\"odinger operator acting on $\ell^2(\zz)$. Let $v_{\omega, L}$ be 
the periodic potential on $\zz$ obtained by repeating the restriction of
$v_\omega$ to $[-L,L]$. In~\cite{La3}, it is  proved that for any interval $I$, 
\beq
\limsup_{L\rightarrow \infty}|\sp_{\ac}(h_{\omega, L})\cap I|\leq |\sp_{\ac}(h_\omega)\cap I|
\label{paris-late}
\eeq
holds with probability one.\footnote{$|\,\cdot\,|$ stands for the Lebesgue 
measure.} Although motivated by the implication $(1)\Rightarrow(4)$ and the 
study of the Thouless conductance, this results is stronger than one needs for 
this purpose.\footnote{It suffices  to show that 
$|\sp_{\ac}(h_\omega)\cap I|=0 \Rightarrow \lim_{L\rightarrow 
\infty}|\sp_{\ac}(h_{\omega, L})\cap I|=0$.} 
Independent of its motivation, the relation~\eqref{paris-late} was shown to 
have  important  consequences for the spectral theory of quasi-periodic 
operators; see~\cite{La3} for details.

In~\cite{GS}, the relation~\eqref{paris-late} was extended to the 
deterministic setting and to higher dimensions. If this extension was 
applicable to half-line
Schr\"odinger operators $h=-\Delta + v$ acting on $\ell^2(\zz_+)$ with 
periodic approximations $h_{{\rm per}, L}=-\Delta + v_{{\rm per}, L}$ acting 
on $\ell^2(\zz)$ and obtained by repeating the restriction of $v$ to $[1,L]$, 
then the implication $(1)\Rightarrow (4)$ in Theorem \ref{main-th} 
would follow.\footnote{In the ergodic case, the homogeneity of the potential 
yields that half-line and full line periodization are equivalent for the 
purpose of the inequality~\eqref{paris-late}. This is {\em not} the case in 
the deterministic setting.}  Surprisingly, it is not known how to adapt the 
arguments of~\cite{GS} to the half-line case.\footnote{We are grateful to 
Fritz Gestezsy and Barry Simon for discussions regarding this point.}
Our proof of the implication $(2)\Rightarrow (4)$ proceeds by adopting the 
deterministic part of the argument in~\cite{La, La3} (see 
Section~\ref{ssec:periodicop}) and by replacing the ergodic part with alternative 
arguments presented in Section~\ref{sec-japan}. These arguments give
\beq
\limsup_{L\rightarrow \infty} |\sp_{\ac}(h_{{\rm per}, L})\cap I|\leq C |\sp_{\ac}(h)\cap I|^{\frac{1}{5}},
\label{no-end}
\eeq
where  $C=5\left(\frac{\pi^2(1+\pi)^4}{4}\right)^{1/5} \simeq 18.7$; see 
Remark at the end of Section~\ref{sec-japan} and~\cite{BJLP2}. The validity of 
the relation 
$\limsup_{L\rightarrow \infty}|\sp_{\ac}(h_{{\rm per}, L})\cap I|
\leq |\sp_{\ac}(h)\cap I|$ 
in the setting of deterministic half-line Schr\"odinger operators remains an open problem.

The paper is organized as follows. In Section~\ref{sec-cond} we review the 
Landauer-B\"uttiker and Thouless conductance formulas. The proof of 
Theorem~\ref{main-th} is given in the remaining sections. We shall prove 
independently the equivalence between (2) and (1), (3), (4): the equivalence 
$(1)\Leftrightarrow(2)$ is proven in Section~\ref{sec-proof12}, 
$(2)\Leftrightarrow(3)$ in Section~\ref{sec-proof23} and  
$(2)\Leftrightarrow(4)$ in Section~\ref{sec-proof24}.

\bigskip\noindent
{\bf Acknowledgment.} The research of V.J. was partly supported by NSERC. The research of Y.L.\ was partly supported by The Israel Science Foundation
(Grant No.\ 1105/10) and by Grant No.\ 2010348 from the United States-Israel
Binational Science Foundation (BSF), Jerusalem, Israel. A part 
of this work has been done during a visit of L.B. to McGill University supported 
by NSERC. Another part was done during the visits of V.J. to The Hebrew University supported by NSERC and to Cergy-Pontoise University supported by the ERC grant DISPEQ. V.J. wishes to thank N. Tzvetkov for making this second visit possible.
The work of C.-A.P. has been carried out in the framework of the Labex Archimède
(ANR-11-LABX-0033) and of the A*MIDEX project (ANR-11-IDEX-0001-02),
funded by the ``Investissements d'Avenir'' French Government programme
managed by the French National Research Agency (ANR).


\section{The Landauer-B\"uttiker and Thouless formulas}
\label{sec-cond}
 
In this section we briefly describe the Landauer-B\"uttiker and Thouless 
conductance formulas  of a finite sample,  referring  the reader 
to~\cite{BJP,BJLP} for a more detailed exposition. The Hilbert space 
describing the sample is $\cH_L=\ell^2(\Z_L)$, where $\Z_L= [1,L]\cap \zz_+$, and its 
Hamiltonian is the discrete Schr\"odinger operator $h_L=-\Delta +v$, 
\begin{equation}\label{samplehamiltonian}
(h_L\psi)(n)= -\psi(n+1)-\psi(n-1)+v(n)\psi(n), \qquad n\in \Z_L,
\end{equation}
with Dirichlet boundary conditions $\psi(0)=\psi(L+1)=0$. 

\subsection{Landauer-B\"uttiker formula}\label{ssec:lbconductance}

To describe  the Landauer-B\"uttiker formula,  we couple the sample at its 
endpoints to two electronic reservoirs. The combined system is 
considered in the independent electron approximation. The left/right reservoir 
is  described by the following ``one electron data'': Hilbert space 
$\cH_{l/r}$, Hamiltonian $h_{l/r}$, and unit vector $\psi_{l/r}$ that  allows 
to couple the reservoir to the sample. The decoupled (one electron) 
Hamiltonian is
$$
h_{0,L}=h_l + h_L +h_r
$$
acting on $\cH=\cH_l\oplus \cH_L\oplus \cH_r$. The junction between the sample 
and the left/right reservoir is described by the tunneling Hamiltonians
$$
h_{T,l}=|\psi_l\ket \bra \delta_1| +|\delta_1\ket\bra \psi_l|  \qquad {\rm and}\qquad   h_{T,r}=|\psi_r\ket \bra \delta_L| +|\delta_L\ket\bra \psi_r|.
$$
The coupled (one electron) Hamiltonian is
$$
h_{\kappa,L}=h_{0,L}+\kappa(h_{T,l}+h_{T,r}), 
$$
where $\kappa\not=0$ is a coupling constant. The left/right reservoir is  
initially at equilibrium at zero temperature  and chemical potential 
$\mu_{l/r}$. We shall assume that $\mu_l<\mu_r$. In the large time limit the 
coupled system approaches a steady state which carries a non-trivial charge 
current. As observed in~\cite{BJP}, for the purpose of discussing transport 
properties of the coupled system one may assume, without loss of generality, 
that $\psi_{l/r}$ is a cyclic vector for $h_{l/r}$. Hence, passing to the 
spectral representation we may assume that $h_{l/r}$ acts as multiplication by 
$E$ on
\[
\cH_{l/r}=L^2(\rr,\d\nu_{l/r}(E)),
\]
where $\nu_{l/r}$ is the spectral measure of $h_{l/r}$ associated to $\psi_{l/r}$. 

The expectation value of the charge current, from the right to the left, in the steady state is given by the Landauer-B\"uttiker formula, see 
e.g.,~\cite{L,BILP,AJPP,CJM,N},
\begin{equation}\label{eq:lbformula}
\cJ_{{\rm LB}}(L, \mu_l, \mu_r) = \int_{\mu_l}^{\mu_r} {\cal D}_{\rm LB}(L, E) \, \d E,
\end{equation}
where $2\pi {\cal D}_{LB}(L, E)$ is the   transmission probability from the 
right to the left reservoir at energy $E$. One can further prove using 
stationary scattering theory\footnote{The scattering matrix $S$ of the pair 
$(h_{\kappa,L},h_{0,L})$, which  by trace class scattering theory is a 
unitary operator on 
$\cH_\ac(h_{0,L})=\Ran 1_{\ac}(h_{0,L})=\Ran 1_\ac(h_l)\oplus\Ran 1_\ac(h_r)$, 
acts as the operator of multiplication by a unitary $2\times 2$ matrix 
$S(L, E)=\left[\begin{matrix} S_{ll}(L, E)&S_{lr}(L, E)\\
S_{rl}(L, E)&S_{rr}(L, E)\end{matrix}\right]$. One then has 
$2\pi{\cal D}_{\rm LB}(L, E)=|S_{lr}(L, E)|^2=|S_{rl}(L, E)|^2$.}
(see~\cite{Y} for the general theory, and~\cite{Lan} for  a simple proof in the present setting) that 
\begin{equation}\label{eq:transmissionproba}
{\cal D}_{\rm LB}(L, E)= 2\pi\kappa^4|\langle\delta_1,(h_{\kappa,L}-E-\i0)^{-1}\delta_L\rangle|^2\, \frac{\d\nu_{l,\ac}}{\d E}(E)  \,\frac{\d\nu_{r,\ac}}{\d E}(E),
\end{equation}
where $\frac{\d\nu_{{l/r},\ac}}{\d E}$ is the density of the absolutely
 continuous part of the spectral measure $\nu_{l/r}$. 
The unitarity of the scattering matrix implies a uniform bound on the
spectral density
\begin{equation}\label{eq:chargecurrentbound}
0\leq  \cD_{\rm LB}(L,E) \leq \frac{1}{2\pi}.
\end{equation}

We denote the essential support of the absolutely continuous spectrum of 
$h_{l/r}$ by 
\beq
\Sigma_{l/r}=\left\{E\ :\ \frac{\d\nu_{l/r,\ac}}{\d E}(E)>0\right\}.
\label{eq:Simgalr}
\eeq
It follows  immediately from (\ref{eq:transmissionproba}) that only energies 
belonging to  $\Sigma_{l}\cap \Sigma_r$ contribute to transport: for any $L$,  
${\cal D}_{\rm LB}(L, E)=0$ whenever $E\notin \Sigma_l\cap \Sigma_r$.  
This leads to the transparency condition mentioned in Remark~1 after 
Theorem~\ref{main-th}, which  is needed for  the proof of implication 
$(3)\Rightarrow(2)$ in Theorem \ref{main-th}:
\begin{definition} The reservoirs are transparent for energies in $(\mu_l, \mu_r)$ if $(\mu_l, \mu_r)\subset 
\Sigma_l\cap \Sigma_r$.
\label{transparent}
\end{definition}

An additional insight into the structure of the EBBM and 
${\cal D}_{\rm LB}(L, E)$ can be obtained by implementing a  spatial structure 
of the reservoirs, see Remark~7 after Theorem~1.1. in~\cite{BJLP}.


\subsection{Thouless formula}\label{ssec:thoulessconductance}

The Thouless formula is the Landauer-B\"uttiker formula of a specific EBBM 
(named the crystalline EBBM in \cite{BJLP}) in which the reservoirs are 
implemented in such a way that  the coupled Hamiltonian  is a periodic 
discrete Schr\"odinger operator on $\ell^2(\zz)$. More precisely, one extends 
the sample potential $v(n)$ to $\zz$ by setting $v(n+ mL)=v(n)$ for 
$n\in \zz_L$ and $m\in \zz$. We denote this extension by $v_{{\rm per},L}$. 
Let  $h_{{\rm per},L}=-\Delta + v_{{\rm per},L}$ be the corresponding periodic 
discrete Schr\"odinger operator acting on 
$\ell^2(\zz)$. 
The Hilbert space ${\cal H}_l$ is $\ell^2((-\infty, 0])$ and the Hilbert space 
$\cH_r$ is  $\ell^2([L+1, \infty))$. The single electron Hamiltonian of the 
left/right reservoir is $h_{{\rm per},L}$ restricted to 
$(-\infty, 0]/[L+1, \infty)$ with Dirichlet boundary condition.  Finally,  
$\psi_l=\delta_{0}$, $\psi_r=\delta_{L+1}$ and $\kappa=1$. The one electron 
Hilbert space of the coupled system is $\ell^2(\zz)$ and the one electron 
Hamiltonian is $h_{{\rm per},L}$. In this case $2\pi {\cal D}_{\rm LB}(L, E)$ 
is the characteristic function of the spectrum of $h_{{\rm per},L}$  and the 
corresponding Landauer-B\"uttiker formula coincides  with the  Thouless 
formula:
\beq
\cJ_{\rm Th}(L, \mu_l,\mu_r) = \frac{1}{2\pi}|\sp(h_{{\rm per},L})\cap (\mu_l, \mu_r)|,
\label{thouless-heuri}
\eeq
and
\beq
G_{\rm Th}(L,\mu_l,\mu_r)= \frac{1}{\mu_r-\mu_l}\cJ_{\rm Th}(L, \mu_l,\mu_r) = \frac{|\sp(h_{{\rm per},L})\cap (\mu_l, \mu_r)|}{2\pi|(\mu_l,\mu_r)|}.
\label{thouless-heuri-2}
\eeq

We refer the reader to \cite{BJLP} for a detailed discussion regarding the 
identification of (\ref{thouless-heuri})  with the usual heuristically 
derived  Thouless conductance formula one finds in the physics literature (see 
also Remark 1 at the beginning of Section \ref{ssec:periodicop} for a short 
explanation).


\section{AC spectrum and transfer matrices}\label{sec-proof12}

In this section we prove the equivalence between $(1)$ and $(2)$. Recall that the spectral measure for the operator $h=-\Delta + v$ on $\ell^2(\zz_+)$ and vector $\delta_1$ is denoted by $\nu_D$. Recall also the definition (\ref{transfer}) of the  transfer matrix of the operator $h$.
We shall often use that $\|T(L,E)\|\geq 1$, which follows directly from 
${\rm det}(T(L,E))=1$.


\subsection{Proof of (1) $\Rightarrow$ (2)}
\label{sec-paris}

The main tool in this section is the following result. Let $u=(1,0)^T$. 
\bet For any $f\in C_0(\rr)$, 
\[
\lim_{L\rightarrow \infty}\frac{1}{\pi}\int_\rr f(E)\|T(L, E)u\|^{-2}\d E=\int_\rr f(E)\d\nu_D(E).
\]
\label{bb}
\eet
This theorem can be traced back to \cite{Ca} in the context of continuous Schr\"odinger operators. In  the discrete case considered here,  it has been proven in \cite{KR, Si3}.


Suppose now that~(1) holds, i.e., that 
$\sp_\ac(h)\cap (\mu_l,\mu_r)=\emptyset$, and let $\epsilon >0$ be given. 
Since $\nu_D\upharpoonright (\mu_l, \mu_r)$ is a singular measure, one can 
find finitely many disjoint open intervals 
$I_1, \cdots, I_\ell$ in $(\mu_l, \mu_r)$ such that $B=\cup_{j=1}^\ell I_j$ 
satisfies 
\[
|B|< \frac{\epsilon}{3}, \qquad 
\nu_D((\mu_l, \mu_r)\setminus B)< \frac{\epsilon}{3\pi}. 
\]
Let $f\in C_0(\rr)$ be a continuous function such that $0\leq f(E)\leq 1$ for all $E$, $f(E)=0$ if and only if $E\in \bar B$, and 
\[
|\{E\in (\mu_l, \mu_r)\,:\, 0 <f(E)<1\}|<\frac{\epsilon}{3}.
\]
Obviously, 
\beq
\int_{\mu_{l}}^{\mu_r} f(E)\d\nu_D(E)<\frac{\epsilon}{3\pi}.
\label{bet}
\eeq
Since $\|T(L, E)\|\geq 1$, the estimate 
\[
\int_{\mu_l}^{\mu_r}\|T(L, E)\|^{-2}\d E 
\leq \int_{\mu_l}^{\mu_r}f(E)\|T(L, E)u\|^{-2}\d E 
+ \int_{\{E\in (\mu_l, \mu_r)\,:\, f(E)<1\}}\|T(L, E)\|^{-2}\d E
\]
gives
\[\int_{\mu_l}^{\mu_r}\|T(L, E)\|^{-2}\d E 
\leq \int_{\mu_l}^{\mu_r}f(E)\|T(L, E)u\|^{-2}\d E + \frac{2\epsilon}{3}.
\]
Theorem~\ref{bb} and the estimate~\eqref{bet} now give
\[
\limsup_{L\rightarrow \infty}\int_{\mu_l}^{\mu_r}\|T(L, E)\|^{-2}\d E 
<\epsilon.
\]
Since $\epsilon>0$ is arbitrary, this proves that~(2) holds true for any sequence $(L_k)$ satisfying $\lim L_k=\infty$.


\subsection{Proof of (2) $\Rightarrow$ (1)}

Let $(L_k)$ be a sequence such that 
\begin{equation}\label{eq:21proof}
\lim_{k\rightarrow \infty}\int_{\mu_l}^{\mu_r}\|T(L_k, E)\|^{-2}\d E =0.
\end{equation}
Since $\|T(L_k,E)\|^{-2}\leq 1$, there exists a subsequence of $(L_k)$, which we denote by the same letters, such that for Lebesgue 
a.e.\;$E\in (\mu_l, \mu_r)$, 
\[
\lim_{k\rightarrow \infty}\|T(L_k, E)\|^{-2}=0.
\]
By the result of Last and Simon (recall Remark~6),
\[
\Sigma_\ac \subset \left\{ E\,:\, \liminf_{k\rightarrow \infty}\|T(L_k, E)\|<\infty\right\},
\]
where the inclusion is modulo a set of Lebesgue measure zero.
Hence, $\nu_\ac([\mu_l, \mu_r])=0$, and we can conclude that 
$\sp_\ac(h)\cap (\mu_l, \mu_r)=\emptyset$.

\section{Transfer matrices and  Landauer conductance}
\label{sec-proof23}

In this section we prove the equivalence between $(3)$ and $(2)$. 
Our main tool  is the following result which is an immediate consequence of 
Theorem~1.3 in~\cite{BJP}.
\bet\label{cold}
Let $(L_k)$ be any sequence of positive integers such that 
$\lim L_k=\infty$. Then 
\[
\left\{E\in \Sigma_l\cap \Sigma_r\,:\, \lim_{k\rightarrow \infty} 
{\cal D}_{\rm LB}(L_k, E)=0\right\}=\left\{E\in \Sigma_l\cap \Sigma_r\,:\,
 \lim_{k\rightarrow \infty} 
\|T(L_k, E)\|=\infty\right\}
\]
where the equality is modulo a  set of Lebesgue measure zero.
\eet
This  theorem yields:
\bep\label{cold-1}
Let $I\subset \rr$ be a bounded interval and $(L_k)$ a sequence of positive 
integers such that $\lim L_k=\infty$. Then the following statements are 
equivalent:
\begin{enumerate}[{(i)}]
\item $$\lim_{k\rightarrow \infty}\int_I {\cal D}_{\rm LB}(L_k, E)\d E=0.$$
\item $$
\lim_{k\rightarrow \infty}\int_{I\cap \Sigma_l\cap \Sigma_r} \|T(L_k, E)\|^{-2}\d E=0.$$
\end{enumerate}
\eep
{\bf Remark.} Note that  $\int_I {\cal D}_{\rm LB}(L_k, E)\d E=\int_{I\cap \Sigma_l\cap \Sigma_r} {\cal D}_{\rm LB}(L_k, E)\d E$.

\proof We will prove the implication {\it (i)}$\Rightarrow${\it (ii)}. The 
proof of the reverse implication is identical. 

We argue by contradiction. Suppose that {\it (i)} holds and {\it (ii)} fails. 
Take a subsequence of $(L_k)$, which we denote by same letters, such that 
\beq
\lim_{k\rightarrow \infty}\int_{I\cap \Sigma_l\cap \Sigma_r} 
\|T(L_k, E)\|^{-2}\d E>0.
\label{freezing}
\eeq
It follows from {\it (i)} and the bound~\eqref{eq:chargecurrentbound} that 
there is a subsequence of $(L_k)$, which we denote by the same 
letters, such that for Lebesgue a.e.\;$E\in I\cap \Sigma_l\cap \Sigma_r$, 
$\lim_{k\rightarrow \infty}{\cal D}_{\rm LB}(L_{k}, E)=0$. 
Theorem~\ref{cold} and dominated convergence then give 
$\lim_{k\rightarrow \infty}\int_{I\cap \Sigma_l\cap \Sigma_r} \|T(L_k, E)\|^{-2}\d E=0$, 
contradicting~\eqref{freezing}.\hfill\qed

Returning to Theorem~\ref{main-th}, Proposition~\ref{cold-1} yields the 
implication $(2)\Rightarrow(3)$. If in addition the reservoirs are transparent 
for energies in the interval $I=(\mu_l, \mu_r)$ (recall 
Definition~\ref{transparent}), this proposition also yields 
$(3)\Rightarrow(2)$. 


\section{Transfer matrices and  Thouless conductance}
\label{sec-proof24}

\subsection{Periodic operators}
\label{ssec:periodicop}

In this section we review  several general properties of periodic 
Schr\"odinger operators which we will use in the next two  sections. Some are 
well known and we will just recall them, referring  the reader to Chapter~5 
of~\cite{Si} for proofs and additional information. For the readers' 
convenience, we shall include the proofs of results which are less standard 
or for which we do not have a convenient reference.
Throughout this  section, $v_{\rm per}$ denotes an $L$-periodic potential on 
$\zz$ and $h_{\rm per}=-\Delta + v_{\rm per}$ acting on $\ell^2(\zz)$. 

For any $k\in \rr$ and $m\in \zz$ let
\[
H(k,m)=\left[
\begin{matrix}
v_{\rm per}(m+1) &  -1   & \cdots &0& -\e^{-\i k L}\\
-1  &  v_{\rm per}(m+2) &  \cdots & 0&0\\
\vdots & \vdots &\ddots & \vdots& \vdots\\
0&0&\cdots&v_{\rm per}(m+L-1)&-1\\
-\e^{\i kL}  &   0    &\cdots &-1& v_{\rm per}(m+L)
\end{matrix}
\right].
\]
and denote by $E_1(k)\le\ldots\le E_L(k)$ the repeated eigenvalues of
$H(k,0)$. The functions $\rr\ni k\mapsto E_\ell(k)$ are 
$2\pi/L$-periodic and even. They are strictly monotone and real analytic on 
the interval $(0,\pi/L)$. Moreover, they satisfy
$$
E_L(0)>E_L(\frac{\pi}L)\ge E_{L-1}(\frac{\pi}{L})
>E_{L-1}(0)\ge E_{L-2}(0)>\cdots
$$
This implies in particular that each $E_\ell(k)$ is a simple eigenvalue of 
$H(k,0)$ for $k\in (0,\pi/L)$. It follows that for each 
$\ell\in\{1,\ldots,L\}$ there is a unique real analytic function
$$
(0,\pi/L)\ni k\mapsto\vec u_\ell(k)
=(u_\ell(k,1),\ldots,u_\ell(k,L))^T\in\cc^L,
$$ 
such that $H(k,0)\vec u_\ell(k)=E_\ell(k)\vec u_\ell(k)$, $u_\ell(k,1)>0$ and
$\|\vec u_\ell(k)\|=1$. A bounded two-sided sequence 
$u_\ell(k)=(u_\ell(k,m))_{m\in\zz}$ is obtained by setting
\beq
u_\ell(k,j+nL)=\e^{\i knL}u_\ell(k,j),
\label{eq:eigenvectors}
\eeq
for any $j\in\{1,\ldots,L\}$ and $n\in\zz$. Then, for any $m\in\zz$,
$$
\vec u_\ell(k,m)=(u_\ell(k,m+1),\ldots,u_\ell(k,m+L))^T,
$$
is a normalized eigenvector of $H(k,m)$ for the eigenvalue $E_\ell(k)$.

It follows from Floquet theory that $E\in\sp(h_{\rm per})$ iff the eigenvalue 
equation
\beq\label{eq:stationaryschrodinger}
h_{\rm per} u = Eu
\eeq
has a non-trivial solution $u$ satisfying $u(n+L)=\e^{\i kL}u(n)$ for some 
$k\in\rr$ and all $n\in\zz$. This solution is called Bloch wave of energy
$E$ and $u$ is such a Bloch wave if and only if $E=E_\ell(k)$ for some $\ell$ 
and $(u(m+1),\ldots,u(m+L))^T$ is an eigenvector of $H(k,m)$ for $E_\ell(k)$.
In particular, for any $m$,
$$
\sp(h_{\rm per})=\bigcup_{k\in [0,\pi/L]}\sp(H(k,m))=\bigcup_{\ell=1}^L B_\ell,
$$
where $B_\ell$ is the closed interval with boundary points $E_\ell(0)$ and 
$E_\ell(\pi/L)$. The $B_\ell$ are called spectral bands of $h_{\rm per}$
and have pairwise disjoint interiors. $E$ is an interior point of $B_\ell$ 
iff $E=E_\ell(k)$ for some $k\in(0,\pi/L)$. Moreover, $u$ is a Bloch wave of
energy $E$ iff $u(n)=c u_\ell(k,n)$ for some non-vanishing $c\in\cc$. The
integer $\ell$ is called the band index and number $k$ the quasi-momentum of 
$u$. We say that $u$ is normalized if $|c|=1$.

\noindent {\bf Remark 1.} Here one can see the origin of the mathematical 
definition~\eqref{thouless-heuri-2} of Thouless conductance. Thouless 
conductance associated to an interval $I$ was initially defined (see, e.g., 
\cite{ET}) as the ratio $\frac{\delta E}{\Delta E}$ where $\delta E$ is the 
energy uncertainty within the window $I$ due to a change of boundary condition 
and $\Delta E$ is the mean level spacing in $I$. 
The energy uncertainty within a single energy band $B_\ell\subset I$ is of the 
order of the band width $|B_\ell|=|E_\ell(\pi/L)-E_\ell(0)|$ which coincides 
with the variation of the eigenvalue $E_\ell(k)$ as the Bloch boundary 
condition changes from periodic to anti-periodic. Convenient estimates for 
$\delta E$ and $\Delta E$ are then given by
$$
\delta E\sim\frac{\sum_{B_\ell\subset I}|B_\ell|}{\sum_{B_\ell\subset I}1} \sim\frac{|\sp(h_{\rm per})\cap I|}{\sum_{B_\ell\subset I}1}
\quad \mbox{ and } \quad 
\Delta E\sim\frac{|I|}{\sum_{B_\ell\subset I}1},
$$
and the  Thouless conductance becomes $\ds \frac{\delta E}{\Delta E}\sim\frac{|\sp(h_{\rm per})\cap I|}{|I|},$
which, up to a factor $2\pi$, is precisely~\eqref{thouless-heuri-2}.

\bigskip
The discriminant of $h_\per$ is  $D(E)= \tr(T(L,E))$,  where $T(L,E)$ is the 
transfer matrix over one period. The characteristic polynomial of $H(k,m)$ 
satisfies
$$
\det(H(k,m)-z)=D(z)-2\cos(kL).
$$
As a consequence, $\sp(h_{\rm per})= D^{-1}([-2,2])$ and on each band $B_\ell$ 
of $\sp(h_{\rm per})$ the function $D$ is either strictly increasing or 
strictly decreasing~\cite{Si}. Since ${\rm det}(T(L,E))=1$ one also gets that 
$E\in \sp(h_{\rm per})$ if and only if the matrix $T(L,E)$ has two eigenvalues 
of modulus $1$. They are complex conjugate when $k\in(0,\pi/L)$, i.e., when 
$E$ is in the interior of the bands.

The following lemma was proven in~\cite{La2}.
\begin{lemma}\label{lem:derivativeEk} 
For any $\ell\in\{1,\ldots,L\}$, $k\in(0,\pi/L)$ and $m\in\zz$, the following
holds,
\beq\label{eq:derivativeEk}
E_\ell'(k)=2L\,\Im\left(\overline{u_\ell(k,m)}u_\ell(k,m+1)\right).
\eeq
\end{lemma}

\proof For any $k$ and $m$ the vector 
$\vec u_\ell(k,m)=(u_\ell(k,m+1),\ldots,u_\ell(k,m+L))^T$ is a normalized 
eigenvector of $H(k,m)$ for $E_\ell(k)$. The Feynman-Hellmann formula gives
\begin{align*}
E_\ell'(k)&=\left\langle\vec u_\ell(k,m),
\frac{\d H(k,m)}{\d k}\vec u_\ell(k,m)\right\rangle\\
&=\i L\left(\overline{u_\ell(k,m+1)}\e^{-\i kL} u_\ell(k,m+L)
-\overline{u_\ell(k,m+L)}\e^{\i kL}u_\ell(k,m+1)\right),
\end{align*}
and the relation~\eqref{eq:eigenvectors} yields the result.\hfill\qed

From this lemma we obtain first a general estimate on the size of a given band $B_\ell$ and then a bound on the norm of the transfer matrix $T(L,E)$  in terms of normalized Bloch waves and for $E\in\sp(h_{\rm per})$.
\begin{proposition}\label{prop:generalbandbound} 
For any $\ell\in\{1,\ldots,L\}$, one has $|B_\ell|\leq\frac{2\pi}{L}$.
\end{proposition}

\noindent {\bf Remark 1.} This general estimate on $|B_\ell|$ is not new. It 
has been proven, e.g., in~\cite{BLS} from~\eqref{eq:bandbound} using the 
Deift-Simon estimate, see Theorem~\ref{thm:deiftsimon} and 
Eq.~\eqref{eq:derivativeidentity}. Refinements of this estimate can be found 
in~\cite{ShSo}. We provide here an elementary proof 
using~\eqref{eq:derivativeEk}.

\proof Since $E_\ell(k)$ is a strictly monotone function of $k$ on 
the interval $(0,\pi/L)$ we have
\beq\label{eq:bandbound}
|B_\ell|= \int_0^{\pi/L} \left|E_\ell'(k)\right| \d k.
\eeq
Since \eqref{eq:derivativeEk} holds for any $m\in\zz$, we can write
$$
E_\ell'(k)=\sum_{m=1}^L 2\,\Im\left(\overline{u_\ell(k,m)}u_\ell(k,m+1)\right).
$$
The normalization of $u_\ell$ yields
$$
|E_\ell'(k)|
\leq\sum_{m=1}^L\left(|u_\ell(k,m)|^2 + |u_\ell(k,m+1)|^2\right)
=\|\vec u_\ell(k,0)\|^2+\|\vec u_\ell(k,1)\|^2=2,
$$
and the result follows.\hfill\qed

The next two  results provide bounds on the norm of the transfer matrix 
$T(L,E)$ for energies $E$ in and out of the spectrum of $h_{\rm per}$. They 
will be of crucial importance in the  proofs  of the equivalence 
$(2)\Leftrightarrow(4)$. The first result, Lemma~3.1 in~\cite{La3},  
concerns energies inside the spectrum.

\begin{proposition}\label{prop:transfermatrixupperbound}
For any $\ell\in\{1,\ldots,L\}$ and $k\in(0,\pi/L)$ one has
$$
\|T(L,E_\ell(k))\|
\leq 2L\left(|u_\ell(k,1)|^2+|u_\ell(k,2)|^2 \right)|E_\ell'(k)|^{-1}.
$$
\end{proposition}

\proof Since $E_\ell(k)$ is in the interior of a spectral band, the
transfer matrix $T(L,E_\ell(k))$ has two complex conjugate eigenvalues 
$\e^{\pm\i kL}$. It is  easy to see from~\eqref{eq:stationaryschrodinger} 
and the definition of the transfer matrix that
\beq\label{eq:transfereigenvectors}
\vec x_+
=\left[\begin{matrix} x_2\\ x_1\end{matrix}\right]
=\left[\begin{matrix} u_\ell(k,2)  \\ u_\ell(k,1) \end{matrix} \right]
\quad \mbox{ and } \quad  
\vec x_- = \left[\begin{matrix} \bar x_2\\ \bar x_1 \end{matrix} \right]
\eeq
are associated eigenvectors. In particular 
$\|\vec x_+\|^2=\|\vec x_-\|^2 = |u_\ell(k,1)|^2+|u_\ell(k,2)|^2$.

Let $a,b\in\cc$ such that $|a|^2+|b|^2=1$. For $\vec y=a\vec x_++b\vec x_-$ 
one has
\begin{eqnarray*}
\frac{\|T(L,E) \vec y\|^2}{\|\vec y\|^2} & = & \frac{\|a\e^{\i kL}\vec x_+ + b\e^{-\i kL}\vec x_-  \|^2}{\|a\vec x_+ +b\vec x_- \|^2} \\
 & \leq & \frac{\|\vec x_+\|^2(|a|+|b|)^2}{|a|^2\|\vec x_+\|^2+|b|^2\|\vec x_-\|^2-2|a||b| |\langle \vec x_+,\vec x_-\rangle|} \\
 & \leq & \frac{2\|\vec x_+\|^2}{\|\vec x_+\|^2 - |\langle \vec x_+,\vec x_-\rangle| }\\
 & \leq & \frac{2\|\vec x_+\|^2 \left(\|\vec x_+\|^2+|\langle \vec x_+,\vec x_-\rangle|\right)}{\|\vec x_+\|^4 - |\langle \vec x_+,\vec x_-\rangle|^2}\\
 & \leq & \frac{4\|\vec x_+\|^4}{\|\vec x_+\|^4 - |\langle \vec x_+,\vec x_-\rangle|^2}.
\end{eqnarray*}
Therefore
$$
\|T(L,E)\|^2 \leq \frac{4\|\vec x_+\|^4}{\|\vec x_+\|^4 - |\langle \vec x_+,\vec x_-\rangle|^2}.
$$
Now, a simple computation shows that
\begin{eqnarray*}
\|\vec x_+\|^4 - |\langle \vec x_+,\vec x_-\rangle|^2 = \left(|x_1|^2+|x_2|^2\right)^2 - |x_1^2+x_2^2|^2 = 4(\Im(x_1\overline{x_2}))^2,
\end{eqnarray*}
and hence
$$
\|T(L,E)\| \leq \frac{\|\vec x_+\|^2}{|\Im(x_1\overline{x_2})|}.
$$
The result now follows  from~\eqref{eq:transfereigenvectors} and 
Lemma~\ref{lem:derivativeEk}.
\hfill\qed

The second result, Lemma~5.3 in~\cite{La}, complements 
Proposition~\ref{prop:transfermatrixupperbound} and provides a lower bound on 
the norm of the transfer matrix for energies outside the spectrum of 
$h_{\rm per}$. We recall that $D(E)=\tr(T(L,E))$ denotes the discriminant, 
that $\sp(h_{\rm per})=D^{-1}([-2,2])$ and that $D$ is a strictly monotone 
function of $E$ on each band $B_\ell$ of spectrum.

\begin{figure}
\centering
\includegraphics[scale=0.5]{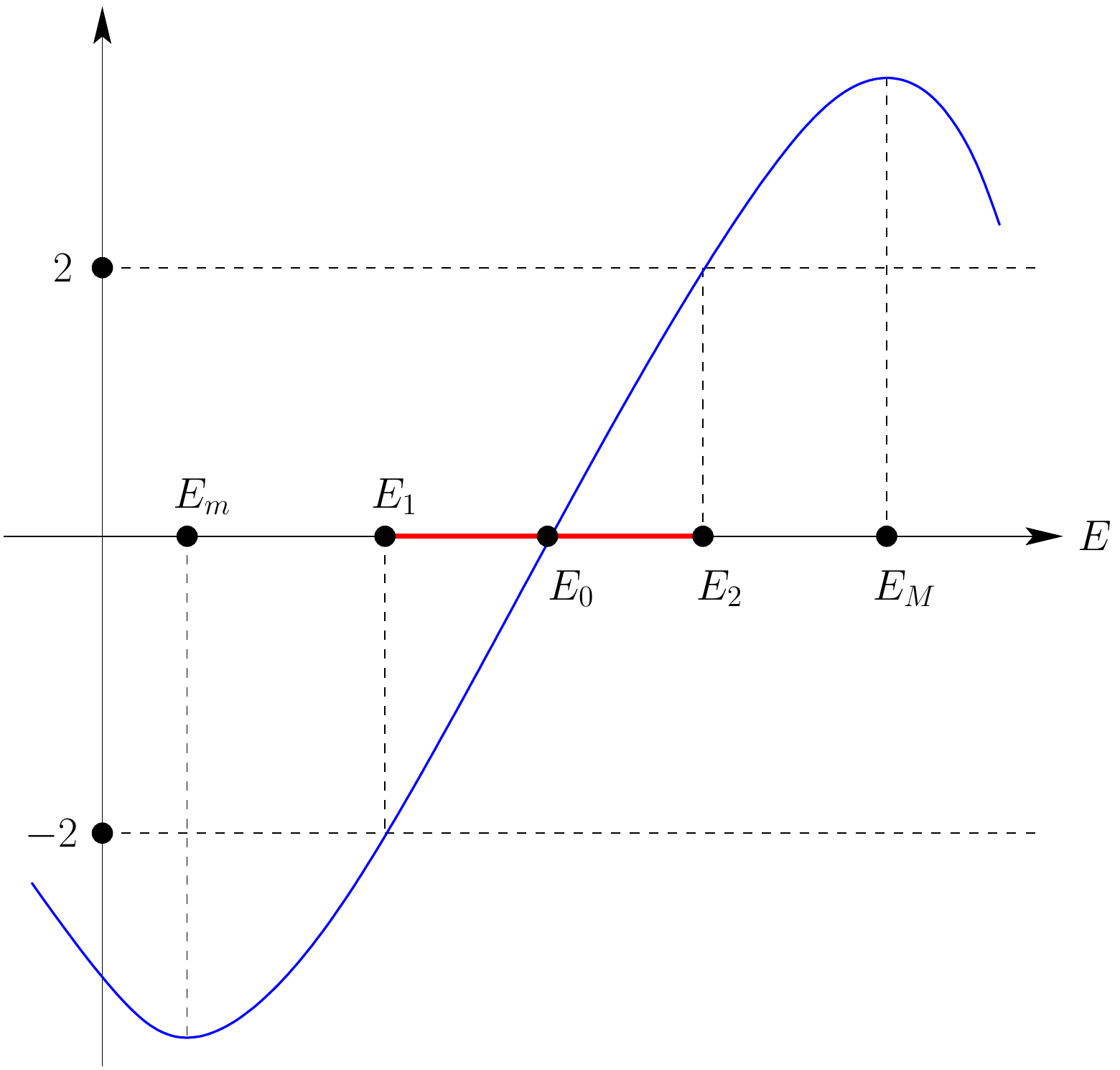}
\caption{The discriminant $D(E)$ near the spectral band $B=[E_1,E_2]$.}
\label{Fig1}
\end{figure}

\begin{proposition}\label{lem:transfermatrixbound} 
Let $B=[E_1,E_2]$ be a spectral band of $h_{\rm per}$. Denote by $E_m$ and $E_M$ the local extrema of $D(E)$ just below and above $B$ (one may be infinite if $B$ is an extremal band) and let $E_0$ be the unique zero of $D(E)$ inside $B$ (see Figure~\ref{Fig1}). Then
\begin{enumerate}[(i)]
 \item For $E\in [E_2,E_M]$, $\ds \|T(L,E)\| \geq \frac{E-E_0}{\e(E_2-E_0)}$.
 \item For $E\in [E_m,E_1]$, $\ds \|T(L,E)\| \geq \frac{E_0-E}{\e(E_0-E_1)}$.
\end{enumerate}
\end{proposition}

\proof Without loss of generality we may assume that $D(E)$ is increasing on $B$. We prove {\it (i)}, the proof of {\it (ii)} is similar.

One easily infers from the definition of the transfer matrix that
$D(E)$ is a real monic polynomial of degree $L$ in $-E$. Since it is positive 
on $(E_0, E_M]$ we can write $D(E)=\prod_{j=1}^L |E-\mathcal{E}_j|$ 
where $\mathcal{E}_j=E_0$ for some $j$.
Hence, we have
$$
f(E)=\frac{\d}{\d E}\ln(D(E)) =\sum_{j=1}^L \frac{1}{E-\mathcal{E}_j},
$$
and
$$
f'(E)=-\sum_{j=1}^L \frac{1}{(E-\mathcal{E}_j)^2}\leq-\frac{1}{(E-E_0)^2}.
$$
Since $E_M$ is a zero of $f$,  for every $E\in(E_0,E_M)$ we can write
$$
f(E) = -\int_{E}^{E_M} f'(E')\, \d E' \geq \int_{E}^{E_M} \frac{1}{(E'-E_0)^2}\, \d E' = \frac{1}{E-E_0}-\frac{1}{E_M-E_0}.
$$
Using the fact that $D(E_2)=2$ we get that for $E\in[E_2,E_M]$
$$
\ln\frac{D(E)}{2}=\ln D(E)-\ln D(E_2)
=\int_{E_2}^E f(E')\d E'\geq\ln\frac{E-E_0}{E_2-E_0}-\frac{E-E_2}{E_M-E_0}
\geq\ln\frac{E-E_0}{E_2-E_0}-1,
$$
from which we obtain
$$
\frac{D(E)}{2} \geq \frac{E-E_0}{\e(E_2-E_0)}.
$$
Since $D(E)=\tr(T(L,E))$ one has $\ds \|T(L,E)\|\geq \frac{D(E)}{2}$ which ends the proof.\hfill\qed

\subsection{Proof of $(4)\Rightarrow(2)$}
We start by following the argument of Lemma~5.1 in~\cite{La}. Recall that 
$h_{{\rm per}, L}$ denotes the periodized Hamiltonian of the sample, see 
Section~\ref{ssec:thoulessconductance}.

For any $L$ we denote the bands of $\sigma_L=\sp(h_{{\rm per},L})$ by 
$B_\ell=[E_{1,\ell},E_{2,\ell}]$, $\ell\in\{1,\ldots,L\}$, and by $D(L,E)$ 
the discriminant of $h_{{\rm per},L}$. We denote by $E_{0,\ell}\in B_\ell$ the 
zeros of $D(L,E)$ and by $E_{m,\ell}$, $\ell\in\{0,\ldots,L\}$, its local 
extrema, so that $E_{m,0}=-\infty$, $E_{m,\ell}\in[E_{2,\ell},E_{1,\ell+1}]$ for $\ell\in\{1,\ldots,L-1\}$ and $E_{m,L}=+\infty$.
Finally, assume that $(L_k)$, 
$\mu_l$ and $\mu_r$ are such that (4) holds, i.e.
\beq\label{eq:proof4-2assumption}
G_{\rm Th}(L_k,\mu_l,\mu_r)=
\frac{|\sigma_{L_k}\cap (\mu_l,\mu_r)|}{2\pi (\mu_r-\mu_l)}\to 0 
\quad \mbox{ as } \ k\to\infty.
\eeq

\begin{figure}
\centering
\includegraphics[scale=0.43]{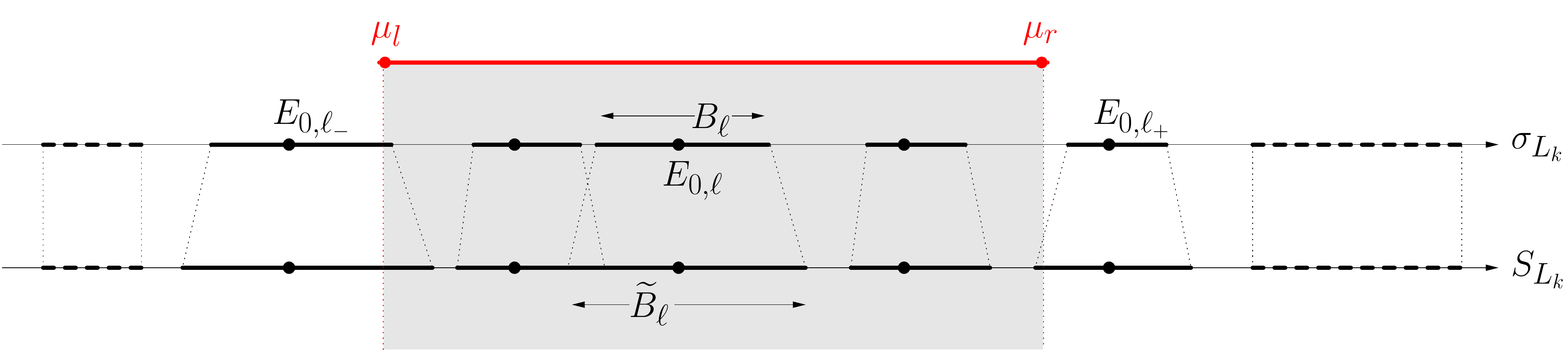}
\caption{The enlarged spectrum $S_{L_k}$.}
\label{Fig2}
\end{figure}

The first step of the proof is to enlarge  $\sigma_{L_k}$  in an appropriate way 
so that energies $E$ which are not in this enlarged spectrum are actually 
``far'' from $\sigma_{L_k}$ (thus, by 
Proposition~\ref{lem:transfermatrixbound}, $\|T(L_k,E)\|$ will be large 
for these energies), while at the same time the measure of enlarged spectrum 
within $I=(\mu_l,\mu_r)$ remains small. The construction goes as follows.  
Let $(c_k)$ be a sequence of positive numbers such that $c_k\to\infty$, $\ds 
\frac{c_k}{L_k}\to 0$ and $c_k|\sigma_{L_k}\cap I|\to 0$. With
$$
\ell_-= \max\{\ell \, : E_{0,\ell} < \inf(I)=\mu_l \} \quad \mbox{ and } \quad 
\ell_+= \min\{\ell \, : E_{0,\ell} > \sup(I)=\mu_r \},
$$
set
$$
\widetilde{B}_\ell=\left\{
\begin{array}{ll}
[E_{0,\ell}-c_k(E_{0,\ell}-E_{1,\ell}),E_{0,\ell}+c_k(E_{2,\ell}-E_{0,\ell})]&
\text{if }\ell_-\leq \ell \leq \ell_+;\\[8pt]
B_\ell&\text{otherwise},
\end{array}
\right.
$$
and define the enlarged spectrum by (see Figure~\ref{Fig2})
\[
S_{L_k}=\bigcup_{\ell=1}^{L_k}\widetilde{B}_\ell.
\]

Note that for $\ell_-<\ell<\ell_+$ one has $E_{0,\ell}\in I$ and a simple
analysis shows that
$$
|\widetilde{B}_\ell\cap I|\leq c_k|B_\ell\cap I|,
$$
while for $\ell=\ell_\pm$, taking Proposition~\ref{prop:generalbandbound} into 
account, we can write
$$
|\widetilde{B}_\ell\cap I|\leq|\widetilde{B}_\ell|=c_k|B_\ell|
\leq\frac{2\pi c_k}{L_k}.
$$
In the other cases, one has
$$
\widetilde{B}_\ell\cap I=B_\ell\cap I=\emptyset.
$$
Thus, the overlap of the extended spectrum with the interval $I$ can be 
estimated as
\begin{align*}
|S_{L_k}\cap I|
\leq\sum_{1\leq\ell\leq L}|\widetilde{B}_\ell\cap I|
&=\sum_{\ell_-<\ell<\ell_+}|\widetilde{B}_\ell\cap I|
+|\widetilde{B}_{\ell_-}\cap I|+|\widetilde{B}_{\ell_+}\cap I|\\
&\leq c_k\sum_{\ell_-<\ell<\ell_+}|B_\ell\cap I|+\frac{4\pi c_k}{L_k}\\
&\leq c_k|\sigma_{L_k}\cap I|+\frac{4\pi c_k}{L_k}.
\end{align*}
Our assumption on the sequence $(c_k)$ ensures that the enlarged spectrum 
still satisfies
$$
|S_{L_k}\cap I|\to 0. 
$$

Suppose now that $E\in I\setminus S_{L_k}$. Then $E\notin\widetilde{B}_\ell$ 
for any $\ell$ and hence must be in one of the intervals 
$(E_{2,\ell},E_{m,\ell}]$ with $E-E_{0,\ell}>c_k(E_{2,\ell}-E_{0,\ell})$ or in 
$[E_{m,\ell},E_{1,\ell+1})$ with 
$E_{0,\ell+1}-E>c_k(E_{0,\ell+1}-E_{1,\ell+1})$. In either case, it follows 
from Proposition~\ref{lem:transfermatrixbound} that 
$$
\|T(L_k,E)\|\geq \frac{c_k}{\e}.
$$
Since $\|T(L_k,E)\|\geq 1$ for any $E$, we derive that, for all $E\in I$ and 
any $k$,
\begin{equation}\label{eq:transfermatrixbound}
\|T(L_k,E)\|\geq  \frac{c_k}{\e}\left(1-\one_{S_{L_k}}(E)\right) + \one_{S_{L_k}}(E),
\end{equation}
where $\one_{S_{L_k}}$ denotes the characteristic function of the set $S_{L_k}$. Hence, for any $k$,
$$
\int_I \|T(L_k,E)\|^{-2} \d E
\leq\left(\frac{\e}{c_k}\right)^2|I\setminus S_{L_k}|+|I\cap S_{L_k}|
\leq\left(\frac{\e}{c_k}\right)^2|I|+|I\cap S_{L_k}|.
$$
The last estimate yields
$$
\lim_{k\to \infty} \int_I \|T(L_k,E)\|^{-2} \d E = 0,
$$ 
and concludes the proof of $(4)\Rightarrow(2)$.

\subsection{Proof of $(2)\Rightarrow(4)$}
\label{sec-japan}

Again in this section $h_{{\rm per}, L}$ denotes the periodized Hamiltonian of 
the sample acting on $\ell^2(\zz)$. The main part of the proof concerns  fixed 
$L$ and we shall occasionally simplify the notation by omitting  the $L$ 
dependence of  various quantities. 

We first introduce some notation. If $E$ is an interior point of the
spectral band $B_\ell$ of $h_{{\rm per},L}$, then there exists a unique $k\in(0,\pi/L)$ such that $E=E_\ell(k)$. We write 
$k(E)$ for this unique $k$. The rotation number is the function defined as
\beq\label{def:rotationnumber}
\alpha(E)
=\int_{-\infty}^{E}\left|k'(\mathcal{E})\right|\d\mathcal{E},
\eeq
where, by convention, we set $k'(E)=0$ when $E$ is not an interior point of
any spectral band. Since $k(E)$ is strictly monotone on each $B_\ell$ one 
easily gets that, for any $\ell$,
$$
\int_{B_\ell} \left|k'(\mathcal{E})\right|\d\mathcal{E}=\frac{\pi}{L}.
$$
Hence, $E\mapsto\alpha(E)$ is strictly increasing on $\sp(h_{{\rm per},L})$
and constant on its complement. Thus, it defines a bijection from 
$\sp(h_{{\rm per},L})$ to $[0,\pi]$\footnote{The function $\pi^{-1}\alpha(E)$ 
is actually the integrated density of states of $h_{{\rm per}, L}$; see 
\cite{DS}}. We shall denote by $E(\alpha):[0,\pi]\to \sp(h_{{\rm per},L})$ its 
inverse and re-parametrize the Bloch waves by defining
$$
u(\alpha,m)= u_\ell(k,m),\quad\mbox{for}\quad \alpha=\alpha(E_\ell(k)).
$$
We note that for any $\ell\in\{1,\ldots,L\}$ and $k\in(0,\pi/L)$ one has
\beq\label{eq:derivativeidentity}
E'(\alpha(E_\ell(k)))
=\frac{1}{\alpha'(E_\ell(k))}=\frac{1}{|k'(E_\ell(k))|}
=\left|E_\ell'(k)\right|.
\eeq
A fundamental result about the rotation number is the following estimate due to Deift and Simon \cite{DS}; see also \cite{ShSo}.
\bet\label{thm:deiftsimon}(\cite{DS}, Theorem 1.4) 
For a.e.\;$E\in\sp(h_{{\rm per},L})$,
$$
2\sin (\alpha(E))\alpha'(E)\geq 1.
$$
\eet
We shall only need a weaker version of it, namely the fact that
\beq
|\alpha^{-1}(\mathcal{A})|\leq 2|\mathcal{A}|,
\label{eq:DSestimate}
\eeq
for any measurable set $\mathcal{A}\subset[0,\pi]$.

We now state and prove two preparatory lemmas.
\bel\label{lem:ualphaform}
\[
\int_0^\pi \left(|u(\alpha,1)|^2 + |u(\alpha,2)|^2\right)\d \alpha =\frac{2\pi}{L}.
\]
\eel

\proof Changing the variable of  integration, we can write
$$
\int_0^\pi\left(|u(\alpha,1)|^2+|u(\alpha,2)|^2\right)\d \alpha
=\sum_{\ell=1}^L\int_0^{\pi/L}\left(|u_\ell(k,1)|^2+|u_\ell(k,2)|^2\right) \alpha'(E_\ell(k))|E_\ell'(k)|\d k,
$$
and Eq.~\eqref{eq:derivativeidentity} allows us to rewrite the
right hand side of the last identity as
$$
\int_0^{\pi/L}\left[\sum_{\ell=1}^L
\left(|u_\ell(k,1)|^2+|u_\ell(k,2)|^2\right)\right]\d k
=\int_0^{\pi/L}\left(\|\vec u_\ell(k,0)\|^2+\|\vec u_\ell(k,1)\|^2\right)
\d k=\frac{2\pi}{L}.
$$
\hfill\qed

\bel\label{lem:energyderivativebound}
\begin{enumerate}[(i)]
\item 
$$
\left|\left\{E\in\sp(h_{{\rm per},L})\,:\,
|u(\alpha(E),1)|^2+|u(\alpha(E),2)|^2>\frac{4\pi}{L\epsilon}\right\}\right|
\leq\epsilon.
$$
\item
$$
\left|\left\{E\in\sp(h_{{\rm per},L})\,:\,
\alpha'(E)>\epsilon^{-1}\right\}\right|\leq\pi\epsilon
$$
\end{enumerate}
\eel
\proof {\it (i)} It follows immediately from Lemma~\ref{lem:ualphaform} that
the set
$$
\mathcal{A}=\left\{\alpha\in[0,\pi]\,:\,
|u(\alpha,1)|^2+|u(\alpha,2)|^2>\frac{4\pi}{L\epsilon}\right\}
$$
is such that $|\mathcal{A}|\le\frac{\epsilon}{2}$. The results thus
follows from the Deift-Simon estimate~\eqref{eq:DSestimate}.

 {\it (ii)} Let
$\mathcal{E}=\left\{E\in\rr\,:\,\alpha'(E)>\epsilon^{-1}\right\}$ and note that
$$
\pi=\int_{-\infty}^\infty\alpha'(E)\d E\geq\frac1\epsilon|\mathcal{E}|.
$$
\hfill\qed

\noindent {\bf Proof of Theorem \ref{main-th}, $(2)\Rightarrow(4)$.}
For a.e.\;$E\in\sp(h_{{\rm per},L})$,
Proposition~\ref{prop:transfermatrixupperbound} and 
Eq.~\eqref{eq:derivativeidentity} yield
$$
\|T(L,E)\|\leq2L(|u(\alpha(E),1)|^2+|u(\alpha(E),2)|^2)\alpha'(E).
$$
Let $\epsilon>0$. It follows from Lemma~\ref{lem:energyderivativebound} that 
there exists $\Omega_\epsilon\subset\sp(h_{{\rm per},L})$ such that
$|\Omega_\epsilon|\le(1+\pi)\epsilon$ and 
\[
|u(\alpha(E),1)|^2+|u(\alpha(E),2)|^2\leq\frac{4\pi}{L\epsilon},
\qquad
\alpha'(E)\leq \frac1\epsilon,
\]
for a.e.\;$E\in\sp(h_{{\rm per},L})\setminus\Omega_\epsilon$. Thus, for 
the same $E$, the estimate
$$
\|T(L,E)\|
\leq2L\,\frac{4\pi}{L\epsilon}\,\frac{1}{\epsilon}=\frac{8\pi}{\epsilon^2},
$$
holds. Hence, for any $L$, one has
\begin{align}
\int_{\mu_l}^{\mu_r}\|T(L, E)\|^{-2}\d E & 
\geq\int_{(\mu_l,\mu_r)\cap(\sp(h_{{\rm per}, L})\setminus \Omega_\epsilon)}
\|T(L, E)\|^{-2}\d E\nonumber\\
&\geq\frac{\epsilon^4}{64\pi^2}\big(
|\sp(h_{{\rm per},L})\cap(\mu_l,\mu_r)|-|\Omega_\epsilon|\big). \label{eq:finalestimate}
\end{align}
Suppose now that $(L_k)$ is such that 
\[
\lim_{k\to \infty}\int_{\mu_l}^{\mu_r}\|T(L_k, E)\|^{-2} \d E=0.
\]
Then (\ref{eq:finalestimate}) gives
\[
\limsup_{k\to \infty} |\sp(h_{{\rm per}, L_k})\cap (\mu_l, \mu_r)|
\leq|\Omega_\epsilon|\leq(1+\pi)\epsilon.
\]
Since this holds for any $\epsilon>0$, we have 
\[
 \lim_{k\to \infty} |\sp(h_{{\rm per}, L_k})\cap (\mu_l, \mu_r)| =0,
\]
and  (4) follows.

\bigskip
{\bf Remark.} The argument of Section \ref{sec-paris} gives that 
\[
\limsup_{L\rightarrow \infty}\int_I \|T(L, E)\|^{-2}\d E\leq |\sp_\ac(h)\cap I|.
\]
Combining this estimate with  (\ref{eq:finalestimate}) one gets
$$
\limsup_{L\rightarrow \infty} |\sp(h_{{\rm per},L}) \cap I| \leq \frac{64\pi^2}{\epsilon^4} |\sp_\ac(h)\cap I| +(1+\pi)\epsilon.
$$
Optimizing over $\epsilon$ one derives  the bound (\ref{no-end}) discussed at the end of Remark 7 after Theorem \ref{main-th}. 
A more refined optimization gives the better constant $C=5 (4\pi^4)^{1/5}\simeq 16.5$. These points will be further discussed  in \cite{BJLP2}.


\end{document}